\documentclass[11pt]{article}

\usepackage{epsfig,graphics,}
\usepackage{amsmath}
\usepackage{latexsym}
\usepackage{amssymb}
\usepackage{tabularx}
\usepackage{color}
\usepackage{verbatim}
\usepackage{url}

\newcommand{\todo}[1]{}

\definecolor{xwcolor}{rgb}{0.0,0.522,1.0}

 \newcommand*{\xw}[1] {}

\definecolor{mdjcolor}{rgb}{0.7,0.0,1.0}

\newcommand*{\mdj}[1] {} 

\definecolor{ltycolor}{rgb}{0.5,0.5,0.0}

\newcommand*{\lty}[1] {} 

\definecolor{swcolor}{rgb}{0.0,0.7,0.1}

\newcommand*{\sw}[1] {}

\newcommand{\YDS}{{\rm YDS}}
\newcommand{\BKP}{{\rm BKP}}
\newcommand{\EDF}{{\rm EDF}}
\newcommand{\AVR}{{\rm AVR}}
\newcommand{\OA}{{\rm OA}}
\newcommand{\qOA}{{\rm qOA}}

\newcommand{\mycomment}[1]{ }

\begin{document} 

\title{An Experimental Comparison of Speed Scaling Algorithms with Deadline Feasibility Constraints}

\author{
Ahmed Abousamra\\
Computer Science Department\\
University of Pittsburgh\\
abousamra@cs.pitt.edu\\
\and
David P. Bunde\\
Computer Science Department\\
Knox College\\
dbunde@knox.edu\\
\and
Kirk Pruhs \thanks{ Supported in part by NSF grants CNS-0325353, CCF-0514058, IIS-0534531, and CCF-0830558, and an IBM Faculty Award.  } \\
Computer Science Department\\
University of Pittsburgh\\
kirk@cs.pitt.edu\\
}

\date{}
\maketitle

\begin{abstract}
We consider the first,
and most well studied, speed scaling problem in the
algorithmic literature:
where the scheduling quality of
service measure is a deadline feasibility constraint,
and where the power
objective is to minimize the total energy used.
Four online algorithms for this problem have been proposed in the
algorithmic literature. 
Based on the best upper bound that can be proved on the competitive ratio,
the ranking of the online algorithms from best
to worst is: $\qOA$, $\OA$, $\AVR$, $\BKP$.
As a test case on the effectiveness of competitive analysis to
predict the best online algorithm,
we report on an experimental ``horse race'' between these algorithms
using instances based on web server traces. 
Our main conclusion is that 
the ranking of our algorithms based on their performance in our experiments is 
identical to the order predicted by
competitive analysis. This ranking holds over a large range of possible
power functions, and even if the power objective is temperature.
\end{abstract}

\section{Introduction}

Energy consumption has become a key issue in the design of microprocessors.
Major chip manufacturers, such as Intel, AMD and IBM,
now produce chips with dynamically scalable speeds,
and produce associated software, such as Intel's SpeedStep and AMD's PowerNow,
that enables an operating system to manage
power by scaling processor speed. Thus the operating system should
have an online {\em speed scaling} policy for setting the speed
of the processor, that ideally should work in tandem with a 
{\em job selection} policy for determining which job to run.
In order to be implementable in a real system, these policies must be online since the
system will not in general be aware of which jobs will arrive in the future.

The resulting online optimization problems, generally called speed scaling
problems, have dual objectives as one both
wants to optimize some schedule quality of service objective
and some power related objective.
In this paper we consider the first~\cite{YDS}, 
and most well studied~\cite{YDS,BBCP,CCL+,BKPSTACS,BKP,LY,LLY,AMS,KK}, speed scaling problem in the
algorithmic literature:
where the scheduling quality of
service measure is a deadline feasibility constraint
(each job must be completed by its deadline), and where the power
objective is to minimize the total energy used.

This problem can be more formally described as follows.
A problem instance consists of $n$ tasks. Task $i$ has a release time
$r_i$, a deadline $d_i > r_i$, and work $w_i > 0$.
An online scheduler learns
about a task only at its release time; at this time, the scheduler also learns
the exact work requirement and the deadline of the task.
A schedule specifies for each time a task to be run and a speed at which to
run the task.
The speed is the amount of work performed on the task per unit time.
Thus, a task with work $w$ run at a constant speed $s$ takes time $\frac{w}{s}$
to complete.
More generally, the work done on a task during a
time period is the integral over that time period of the speed at which
the task is run.
A schedule is {\em feasible} if for each task $i$, work
at least $w_i$ is done on task $i$ during $[r_i, d_i]$.
Note that the times at which work is performed on task $i$ do not have to
be contiguous.
Essentially all of the algorithmic literature has assumed a power function,
which specifies the power $P$ usage as a function of the speed $s$ of
the processor, as $P=s^\alpha$, where $\alpha > 1$ is some constant.
Of particular interest is $\alpha=3$ since dynamic power in 
CMOS based processors is approximately the speed cubed.
The energy used during a time period is the integral of the power
over that time period.

It is easy to see that without loss of generality one can adopt Earliest Deadline First ($\EDF$)
as the job selection algorithm. So the problem reduces to finding algorithms for speed scaling.
Four online speed scaling algorithms for this problem has been proposed in the literature.
Table \ref{table:CR} summarizes where each of these algorithms were proposed, and the
best known bounds on the competitive ratio for these algorithms. We now briefly describe
these algorithms:

\smallskip
\noindent
{\bf Average Rate (\boldmath$\AVR$\unboldmath)}
runs each job at a constant speed between its release and its deadline.
The attraction of the algorithm $\AVR$ is that it is in some sense fair
to all jobs.

\smallskip
\noindent
{\bf Optimal Available (\boldmath$\OA$\unboldmath)}
runs at the speed that would be optimal, given the current state, and given that no more tasks will arrive.
This speed can be determined using the offline greedy 
algorithm $\YDS$ from \cite{YDS} for computing an optimal schedule.

\smallskip
\noindent
{\bf \boldmath$\BKP$\unboldmath} 
intuitively computes the least possible speed
that optimal offline schedule $\YDS$ might currently be running at given the
tasks that have arrived to date, and 
then runs at $e$ times that speed. (If the algorithm ran at some constant $q< e$ times this lower bound,
the deadline of some jobs may be missed.)

\smallskip
\noindent
{\bf \boldmath$\qOA$\unboldmath}
runs at
speed equal to some constant $q\ge1$ times the speed that $\OA$
would run in the current state.

\smallskip

\begin{table}[h]
\centering
\begin{tabular}{|c||c|c|}
\hline
Algorithm&\multicolumn{2}{|c|}{General $\alpha$} \\
\hline
&Upper&Lower\\
\hline
General&&$e^{\alpha-1}/\alpha\cite{qOA}$\\
\hline
\AVR\cite{YDS}&$2^{\alpha-1} \alpha^\alpha$\cite{YDS,BBCP}&$(2-\delta)^{\alpha-1} \alpha^\alpha$\cite{BBCP}\\
\hline
\OA\cite{YDS}&$\alpha^\alpha$\cite{BKP}&$\alpha^\alpha$\cite{YDS}\\
\hline
\BKP\cite{BKP}&$2 (\alpha/(\alpha-1))^\alpha e^\alpha$\cite{BKP}&\\
\hline
\qOA\cite{qOA}&$4^\alpha/(2 \sqrt{e \alpha})$\cite{qOA}&$\frac{1}{2q\alpha}4^\alpha(1-\frac{2}{\alpha})^{\alpha/2}$\cite{qOA}\\
&when $q=2 - \frac{1}{\alpha}$& \\
\hline
\hline
&\multicolumn{2}{|c|}{$\alpha=3$} \\
\hline
&Upper&Lower\\
\hline
General&&2.4\\
\hline
\AVR&108&48\\
\hline
\OA&27&27\\
\hline
\BKP&135.6&\\
\hline
\qOA&6.7&\\
&when $q=1.54$&\\
\hline
\end{tabular}
\caption{Where the online algorithms in the literature were proposed, and the best known bounds on the competitive ratios}
\label{table:CR}
\end{table}

Competitive analysis for online scheduling problems in particular, and online problems in general,
is sometimes criticized for a variety of reasons. The most common criticism is that competitive
analysis focuses on worst-case performance, and thus may not predict
the algorithm that performs best in practice, or on average. 
However, competitive analysis likely will not go
away because it can be tractably applied to such a wide range of problems, 
for which it is
not clear how to obtain a useful average case analysis.
Competitive analysis has applied to several reasonable 
algorithms for this problem, and
the search
for optimally competitive algorithms has lead to candidate algorithms
that
would not likely have been discovered by local search and
experimentation.
Plausibly any of these candidate algorithms might be the best experimentally.

So as a test case on the effectiveness of competitive analysis to
predict the best experimental online algorithm,
we report on an experimental ``horse race'' between these speed scaling algorithms.
Our data was based on web traces, which naturally gave release times
and sizes for each job.
We consider several natural ways of adding deadlines
to the jobs, and tweak the data to produce inputs with workloads with 
different levels of spikiness. 
Based on the best upper bound that can be proved on the competitive ratio,
when $\alpha$ is around 3, the ranking of the online algorithms from best
to worst is: $\qOA$, $\OA$, $\AVR$, $\BKP$.
Our experimental results are essentially that over the wide range of input instances we
tried, the order of the algorithms from best to worst was
exactly the same order as predicted by the best known upper bounds on the competitive ratio.
Further, the differences between the various algorithms was
significant.
So these experimental results can be viewed as a victory for competitive analysis (or alternatively
as a defeat for critics of competitive analysis).

A priori we intuitively expected $\BKP$ to be 
the best experimental algorithm, not the worst. 
We believed that the reason that the best known competitive ratio
for $\BKP$ was so high was that its non-local nature made it more difficult to analyze accurately.
To understand the conceptual difference between $\BKP$ and $\qOA$, consider
a situation where the current load (unfinished work) is low, but the load in the recent past was high.
In this situation $\BKP$ may run at a high speed, while $\qOA$ definitely will not run at a high speed.
It seemed to us that $\BKP$'s use of the historical load should give it an advantage.
Further, in the extreme, when $\alpha=\infty$,
the energy optimal schedule is one that is optimal with respect to the
maximum speed that it reaches. \cite{BKP} show that $\BKP$ is optimally $e$-competitive
with respect to maximum speed. 
This led us to believe that $\BKP$ is near optimally competitive for large $\alpha$.
Further, there appears to be no obvious reason why the relative performance of the algorithms
should depend on $\alpha$. 
Thus, we expected that $\BKP$ would also be the best algorithm when
the cube-root rule ($\alpha=3$) holds.

Some other experimental observations that we believe are interesting are:
\begin{itemize}

\item 
The performance of $\qOA$ is not so sensitive to the value
of $q$. Picking $q$ to be in the range $[1.5, 2]$, as suggested
by the competitiveness results, gives performance reasonably close to the optimal $q$
for each particular instance.
We select $q=1.5$ in comparison with other algorithms because this is the
value of $q$ suggested by the competitive analysis of $\qOA$.

\item The schedule produced by $\qOA$  uses
less energy than the schedules produced by $\AVR$ or $\BKP$.

\item 
There are two alternative formulations of the algorithm $\BKP$ given
in \cite{BKP}. We find that the one that produces a better (higher) lower bound for
the speed of the optimal algorithm $\YDS$ at the current time, is the worse performing
of the two alternatives.

\item 
The value of $q$ is generally higher for moderately spiky workloads than for flat
ones.  
For flat and moderately spiky workloads, the optimal $q$ for the $\qOA$ algorithm is
usually high--- typically around 4 or higher.
This is significant because it shows that $\BKP$ loses to $\qOA$ even when the multiplier $q$ is relatively
large (and bigger than the multiplier $e$ used in $\BKP$). Intuitively this suggests that the main
reason that $\BKP$ loses relative to $\qOA$ is because of its consideration of load in the recent past.
Further, for these workloads, the optimal value of $q$ tends to increase as $\alpha$ is increased. 

\mycomment{
\item As $\alpha$ is increased, optimal $q$ of $\qOA$ usually increases
  for flat and moderately spiky workloads. Optimal $q$ is usually high
  - around 3 or higher - for both types of workload, and hence it is
  logical that optimal $q$ does not go down, because using a smaller
  $q$ means that jobs preceding a spike would run slower, thus taking
  longer and forcing later jobs - which constitute a spike - to run
  much faster to be able to meet their deadlines, which consequently
  cause the workload to consume more energy. However, it is probably
  beneficial to increase $q$ to allow the jobs preceding a spike to
  finish faster, thus giving more room, i.e. a longer time span, for
  later jobs to complete, and therefore enabling them to run at lower
  speeds; which can result in overall less energy consumption.
}

\item For highly spiky workloads and workloads with a fixed time span
  for all jobs, the optimal value of $q$ for $\qOA$ is quite low---
  near 1.
  This is true for fixed time span workloads regardless of the length
  of the fixed time span. 
  For these workloads, the optimal value of $q$ usually decreases as
  $\alpha$ is increased.
\mycomment{
\item As $\alpha$ is increased, optimal $q$ of $\qOA$ usually decreases
  for highly spiky and fixed time span workloads. For both types of
  workload optimal $q$ is quite low - near 1 - which means that
  running faster is not beneficial, and with a greater value of
  $\alpha$ even more energy is consumed that it makes sense to use
  smaller values of $q$ to reduce consumed energy.
}

\item
\cite{BKP} also showed that $\BKP$ and $\YDS$ are cooling oblivious, i.e.
they are simultaneously constant-competitive with
respect to temperature for all values of the cooling parameter
assuming that the environment
has a fixed ambient temperature and that the device cools
according to Newton's law of cooling.
This led us to also compare the various algorithms with respect to 
the objective of maximum temperature.
The relative ordering of the algorithms with respect to maximum
temperature is the same as their order with
respect to energy consumption. Further, the energy optimal schedule $\YDS$ is better
than $\qOA$ with respect to maximum temperature.

\end{itemize}
All of the implementations of the speed scaling algorithms, and related programs, such as 
those for generating test instances, can be found at 
{\tt http://www.cs.pitt.edu/$\sim$kirk/SpeedScalingExperiments}.
We expect, or at least hope, that these tools will be useful to future researchers.

It is important to note again that the purpose of this paper was to determine how
the performance of the candidate algorithms predicted by competitive
analysis compared to a generic experimental analysis.
Thus we based our input on web traces instead of program traces because
these were more readily available, and still served our purposes.
We acknowledge that the common abstract model for a processor, which we
use in this paper, is a significant simplification of a real processor, and 
that there
are many significant issues that would  have to be addressed in applying
these algorithms in a real setting. But these lower level implementation 
issues are beyond the scope of
this paper.

The rest of the paper is organized as follows. In section \ref{sec:preliminaries} we give more
formal definitions of the problem and algorithms.
In section \ref{sec:setup} we explain our experimental setup.
In section \ref{sec:results} we give our experimental observations.

\section{Preliminaries}
\label{sec:preliminaries}

Newton's Law of heat conduction states that the rate of cooling is
proportional to the difference in temperature between the object and
its environment.
We assume the environment has a fixed temperature and that temperature
is scaled so that the environmental temperature is zero.
A first-order approximation for the rate of change $T'$ of the
temperature $T$ is then $T'=P-bT$, where $P$ is the
power used at time $t$, and $b$ is a constant.

A schedule is $R$-competitive, or $R$-approximate,
for a particular objective function
if the value of that objective function on the schedule is at most
$R$ times the value of the objective function on an optimal schedule.
An algorithm $A$ 
is $R$-competitive, or has competitive ratio $R$, if $A(I)$ is $R$-competitive
for all instances.

We now more formally define the algorithms that we consider in this paper, along
with related concepts.
The span of a job $i$ is $d_i - r_i$.
We start with the offline speed scaling algorithm 
$\YDS$ proposed in \cite{YDS}.
Let  $w(t_1,t_2)$ denote the work that has release time at least $t_1$ and has
deadline  at most $t_2$. The \emph{intensity} $I(t_1, t_2)$ 
of the time interval $[t_1, t_2]$ is
defined to be $w(t_1,t_2)/(t_2 - t_1)$.

\medskip
\noindent
{\bf Algorithm \boldmath$\YDS$\unboldmath} \cite{YDS}:
The algorithm repeats the following steps until all jobs are scheduled:
\begin{enumerate}
\item
Let $[t_1, t_2]$ be the maximum intensity time interval.
\item
The processor will
run at speed $I(t_1,t_2)$ during $[t_1, t_2]$ and schedule all
the jobs comprising $w(t_1,t_2)$, always running the released, 
unfinished task with the earliest deadline. 
\item
Then the instance is modified as if the times $[t_1, t_2]$ didn't exist.
That is, all deadlines $d_i > t_1$ are reduced to
$\max(t_1, d_i - (t_2  -t_1))$, and all release times $r_i > t_1$
are reduced to $\max(t_1, r_i - (t_2 - t_1))$. 
\end{enumerate}

\medskip
\noindent
{\bf Algorithm \boldmath$\qOA$\unboldmath} \cite{qOA}:
The speed is $q \cdot \max_t w(t)/t$, where
$w(t)$ is the unfinished work that has deadline within the next $t$ 
units of time. Here $q$ is some constant that is at least 1. 
We set $q=1.5$ when we compare $\qOA$ to other algorithms. $\OA$ is
just the algorithm $\qOA$ when $q=1$.

\medskip
\noindent
{\bf Algorithm \boldmath$\AVR$\unboldmath} \cite{YDS}:
The speed is
$\sum_{i \in J(t)} \frac{w_i}{d_i -r_i}$,
where $J(t)$ is the collection of tasks $i$ with $r_i \le t \le d_i$.
\medskip

\medskip
\noindent
{\bf Algorithm \boldmath$\BKP$\unboldmath} \cite{BKP}:
For $t_1 \leq t \leq t_2$, let $w(t,t_1,t_2)$ denote the amount of work that 
has release time at least
$t_1$ and deadline at most $t_2$ and that has already arrived by time $t$.
Let $p(t)$ be defined by:
$$
p(t) = \max_{t_1,t_2} \frac{w(t, t_1,t_2)}{(t_2-t_1)} \qquad \textrm{ such
that $t_1 < t \leq t_2$} $$
Let $v(t)$ be defined by:
$$
v(t) =  \max_{t' > t} \frac{ w(t, e t  -  (e-1) t',t')}{e(t' -t)}
$$
In one variation of $\BKP$ in \cite{BKP}, the speed is $e \cdot v(t)$, and in the other variation, it is
$e \cdot p(t)$. 
Note that $w(t,t_1,t_2)$, $p(t)$ and $v(t)$ may be computed by
an online algorithm at time $t$.
It is easy to see that $v(t) \leq p(t)$.

\section{Experimental Setup}
\label{sec:setup}

We use the trace file {\tt epa-http.txt} from the Internet Traffic
Archive (\url{http://ita.ee.lbl.gov/}) to generate the workloads for
our experiments.
This trace contains about 50,000 http requests received during one day
by the EPA's webserver located at Research Triangle Park, NC.
Each http request has two main pieces of information: its time and
the number of bytes in the response.
Some requests received 0 bytes in response, nearly always
corresponding to a 304 (page not modified since last download) or 404
(page not found) response code.
In these cases, we treat the response size as 50 bytes to approximate
the header; this value is small relative to the responses generated by
other requests. 
We treat each http request as a job whose release time is the same as
the http request time, whose work requirement is the number of
transferred bytes in response to the request, and whose deadline is
generated in different ways to produce workloads with specific
characteristics.

Since our trace file contains a day's worth of http requests, it has a
peak period and slow periods corresponding to high traffic and low
traffic reaching the website, respectively.
Since we are
interested in workloads with different degrees of spikiness, we repeat
the set of jobs we create based on the trace file five times to
simulate having requests of five days.

To allow us to run multiple experiments on this one trace, in each
experiment we only create a job for every 20$^{\textrm{th}}$ http
request in the original trace. This allows us to generate 20 different
workloads where the first one starts with the first http request in
the file, the second workload starts with the second http request, and
so on. 
Each of these workloads contains different jobs, but each spans the
entire day and contains similar variations in request density.
In addition to providing multiple input workloads, splitting the trace
this way also keeps the computation time of our simulator reasonable. In our simulations, the results of the different workloads exhibited similar trends, so we arbitrarily chose the workload generated starting with the sixth job to present its simulation results.

We will now explain how the deadline is generated for each type of
workload. Through these explanations the variable $S$ will stand
for a fixed scaling factor, and $N$ will denote some random number.
We start with the flat and fixed span workloads. These
are the most natural since they correspond to requiring
a response time proportional to the request size, and requiring a fixed
response time for every request, respectively. After that, we
consider the moderately and highly spiky workloads. We include them
because we wanted to compare the algorithms on spikier workloads.

\subsubsection{Flat Workload}
\begin{figure}
\centering
\epsfig{file=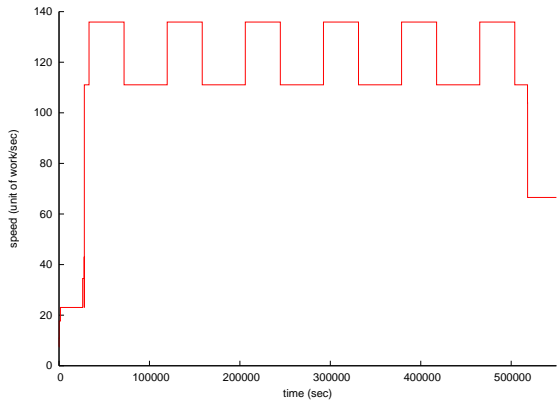, scale=.6}
\caption{Plot of $\YDS$ schedule for the flat workload.}
\label{fig:YDS_Flat}
\end{figure}

The first workload we consider is the {\em flat workload}, in which
the span of each job is proportional to its amount of work.
Although the amount of work varies over time in this workload, it 
does not particularly strain the processor.
We call this work load flat because the optimal energy schedule shows
a relatively modest number and degree of speed changes.
See Figure \ref{fig:YDS_Flat}, which is typical of optimal schedules for
this type of workload.%
\footnote{To make it easy to compare all
  figures, we displayed the same time interval (0--550,000 seconds)
  for all of them, trimming the plots to fit.
  From 550,000 seconds on, the speed continues to decrease as the
  last set of active jobs complete.}

We generate the deadlines of jobs in a flat workload using the equation
$d_i = r_i + Sw_i$, with $S = 0.4$.

\subsubsection{Fixed Span Workload}
\begin{figure}
\centering
\epsfig{file=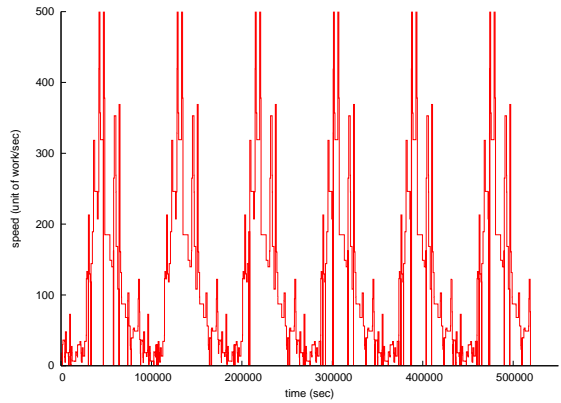, scale=.6}
\caption{Plot of $\YDS$ schedule for the fixed span workload.}
\label{fig:YDS_Fixed}
\end{figure}

Our second workload is the {\em fixed span workload}, in which all jobs
have the same span, corresponding to a system that guarantees a
worst-case response time for each task.
Since jobs vary in their work requirement, the amount of work per unit
time varies.
The optimal schedule produced by $\YDS$ for this kind of workload may
have large or small variations in speed depending on the job span
and how much jobs overlap.
Figure \ref{fig:YDS_Fixed} plots the $\YDS$ schedule for
one of these workloads generated using a fixed span of 1000 seconds.

\subsubsection{Moderately Spiky Workload}

\begin{figure}
\centering
\epsfig{file=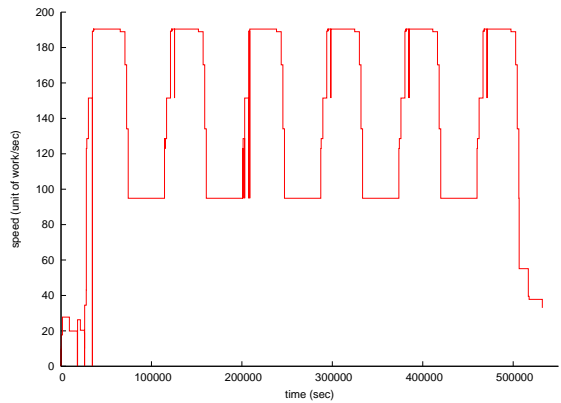, scale=.6}
\caption{Plot of $\YDS$ schedule for the moderately spiky workload.}
\label{fig:YDS_Moderate}
\end{figure}

Our third workload is the {\em moderately spiky workload}, which has
greater variation in the amount of arriving work.
An optimal solution for a moderately spiky workload is shown in 
Figure \ref{fig:YDS_Moderate}.
To generate the deadlines for this workload, we used the equation
$d_i = r_i + Sw_i$, with $S = 0.1$.
Note that this is the same equation we used to generate a flat
workload except for the scaling factor.
The change in the optimal schedule can be seen by comparing
Figures~\ref{fig:YDS_Flat} and \ref{fig:YDS_Moderate}.

\subsubsection{Highly Spiky Workload}

\begin{figure}
\centering
\epsfig{file=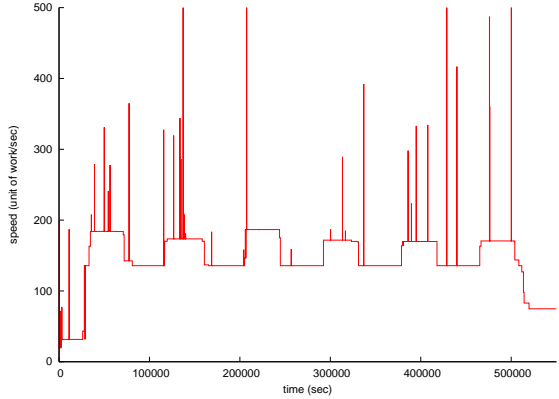, scale=.6}
\caption{Plot of $\YDS$ schedule for the highly spiky workload.}
\label{fig:YDS_High}
\end{figure}

For our last workload, which we call the {\em highly spiky workload}, 
we further increased the variability of the amount of work arriving.
Intuitively, a highly spiky workload contains bursts of high work
when several jobs arrive requiring a lot of work that needs to be
finished in a small time period.
Therefore, a highly spiky workload
contains huge variations in speeds as illustrated by the optimal
$\YDS$ schedule shown in Figure \ref{fig:YDS_High}.

When generating this workload, we added additional jobs as well as
generating job deadlines.
We did this as follows:
\begin{enumerate}
\item We divide the time line into intervals of two alternating
  lengths, $L, H$, with $L > H$. $L$ and $H$ stand for light and high
  load intervals, respectively. We used $L = 200$ and $H = 50$. 
\item For any job, regardless of whether its release time falls in an
  $L$ or $H$ interval, we compute its deadline using the equation:
  $d_i = r_i + Sw_i$, with $S = 0.4$.
\item For a job whose release time falls in an $H$ interval we also
  create zero to two additional jobs with the same release time and
  amount of work. To compute their deadlines, we first compute the
  span of the original job (after computing its deadline in the
  previous step) as $t_i = d_i - r_i$. Then we use the following equation
  for computing the deadline of each additional job:
	  $d_i = r_i + N t_i$, 
  where $N$ is a pseudorandom number selected uniformly over the range
  $(0,2]$.
  To decide how many additional jobs to 
  create, we use a triangle shaped function $f$ over the
  high load interval with peak $=2$, and we compute the            
  value $f(x)$ at $x = r_i$, the release time of the job; the
  number of jobs we generate is then $\lceil f(r_i) \rceil$. 
\end{enumerate}

Note that we use a random number in computing the deadlines of these
jobs so that we do not have multiple identical jobs which would be
equivalent to just one job with the same release time and
deadline and an amount of work equal to the sum of their work.

\section{Experimental Results} 
\label{sec:results}

In this section we show the experimental results for the different
types of workloads. 

\begin{figure}
\centering
\epsfig{file=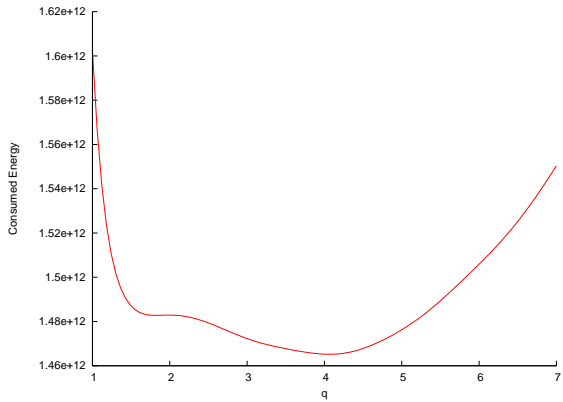, scale=.6}
\caption{Plot of consumed energy vs. $q$ for $\qOA$ schedules of a flat
  workload. ($\alpha=3$)}
\label{fig:q_Flat}
\end{figure}

\begin{figure}
\centering
\epsfig{file=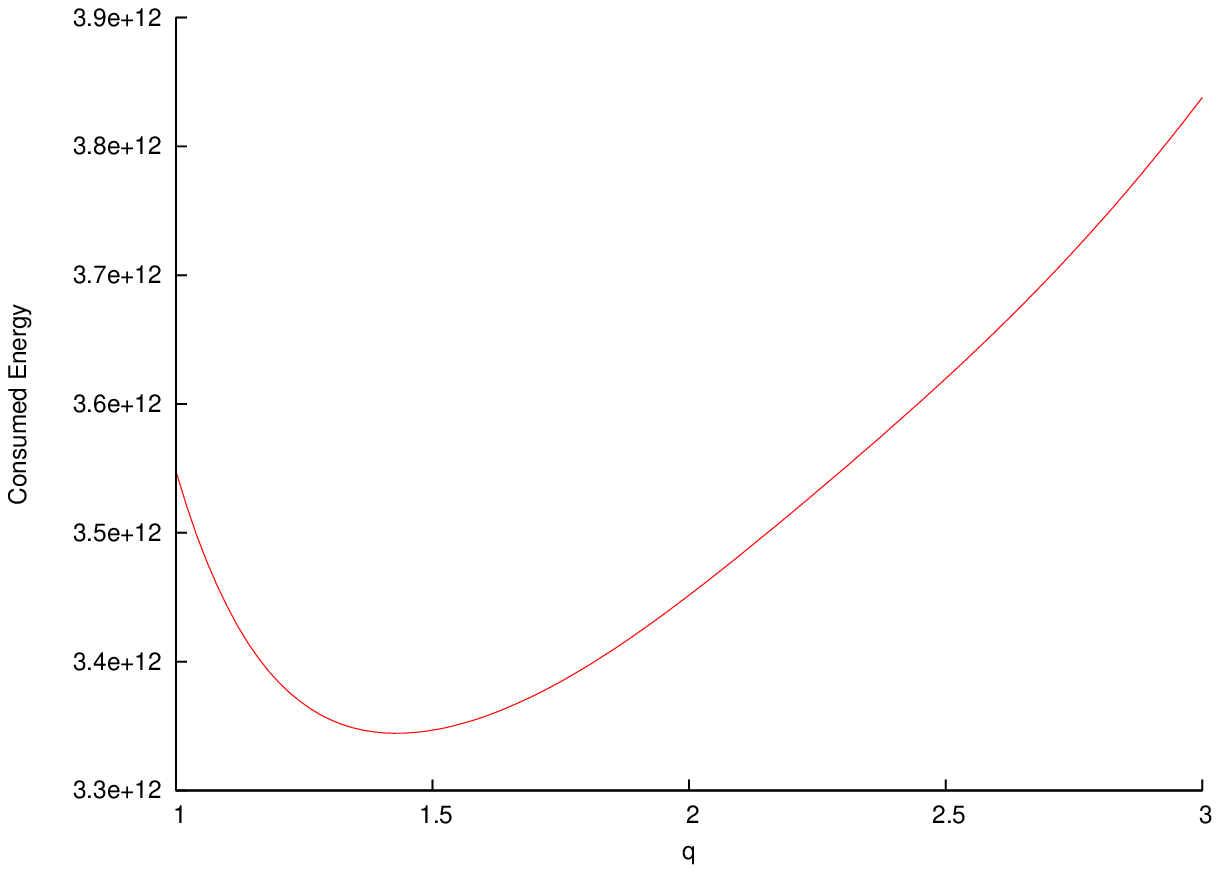, scale=0.6}
\caption{Plot of consumed energy vs. $q$ for $\qOA$ schedules of a
  highly spiky workload. ($\alpha=3$)}
\label{fig:q_High}
\end{figure}

Our first observation addresses one possible concern with using $\qOA$:
how does one pick a good value of $q$?
For each experiment we run $\qOA$ with
different values of $q$ to find the value that results in the least
amount of consumed energy. We tried values of $q$ from $1$ to $9$,
increasing in steps of $0.1$.
It turns out that the performance of $\qOA$ is not highly
sensitive to the exact value of $q$.
Figures \ref{fig:q_Flat} and \ref{fig:q_High} show the consumed
energy as a function of $q$ for $\alpha = 3$ for the flat and
highly spiky workloads, respectively.
The curve for a moderately spiky workload is similar to Figure
\ref{fig:q_Flat}, though less steep to the right of the minimum.
The curve for a fixed time workload is similar to Figure
\ref{fig:q_High}, but with the minimum at 1.
In all cases, the curves are relatively flat near the optimal value of
$q$, implying that any value near the optimal $q$ produces a near
optimal schedule. We set $q=1.5$ when comparing $\qOA$ to other algorithms
because this is the value of $q$ recommended by the competitive analysis of
$\qOA$ when $\alpha$ is about three.

\mycomment{
\begin{figure}
\centering
\epsfig{file=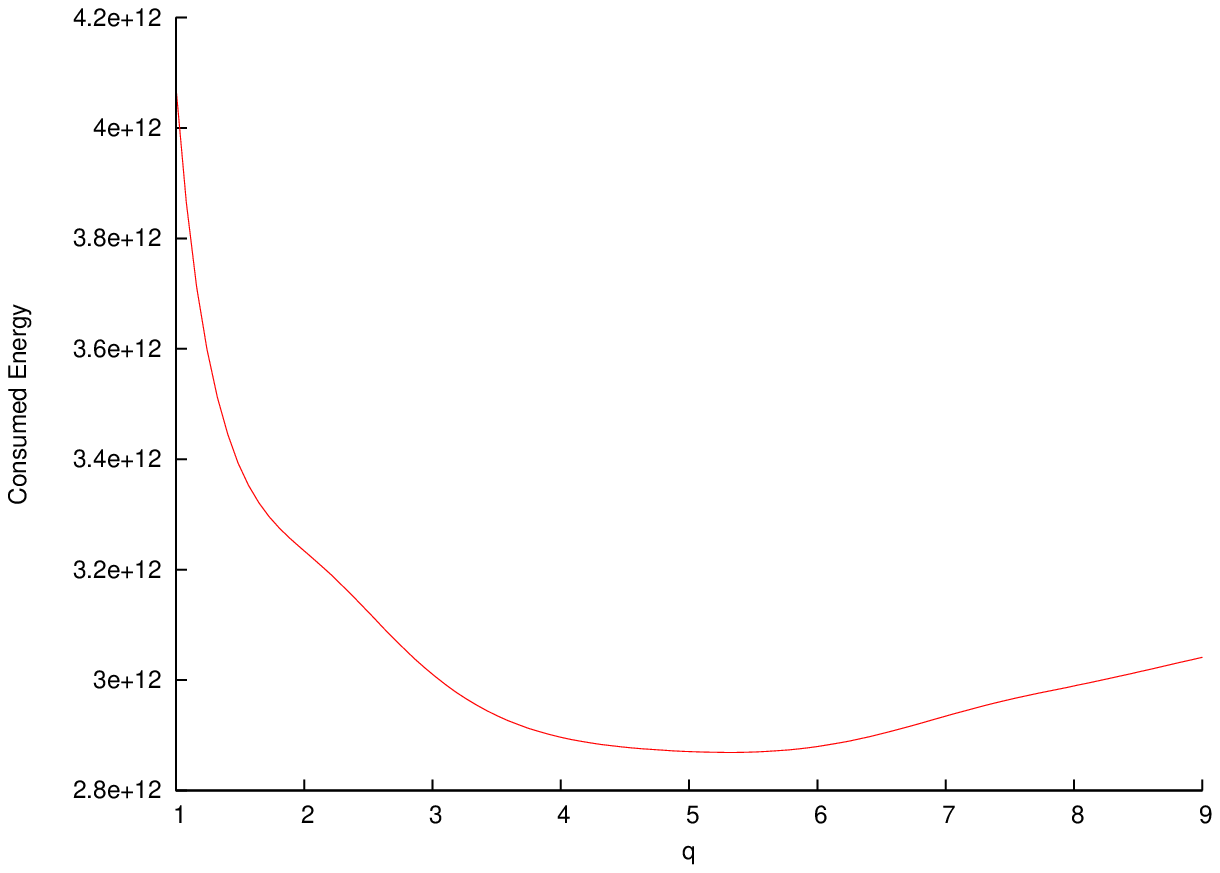, scale=.6}
\caption{Plot of consumed energy vs. $q$ for $\qOA$ schedules of a
  moderately spiky workload. ($\alpha=3$)}
\label{fig:q_Moderate}
\end{figure}
}

\mycomment{
\begin{figure}
\centering
\epsfig{file=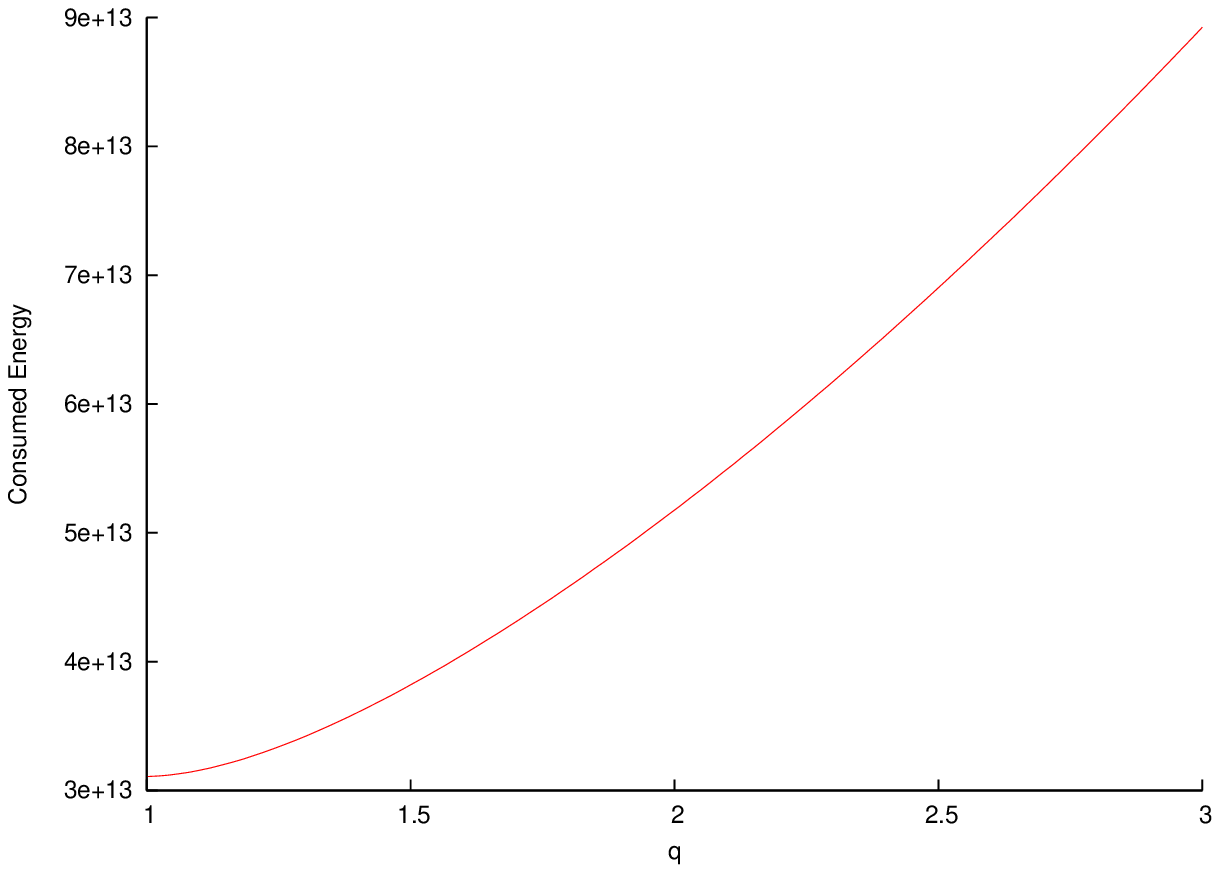, scale=0.6}
\caption{Plot of consumed energy vs. $q$ for $\qOA$ schedules of a
  fixed time workload. ($\alpha=3$)}
\label{fig:q_Fixed}
\end{figure}
}

\begin{figure}
\centering
\epsfig{file=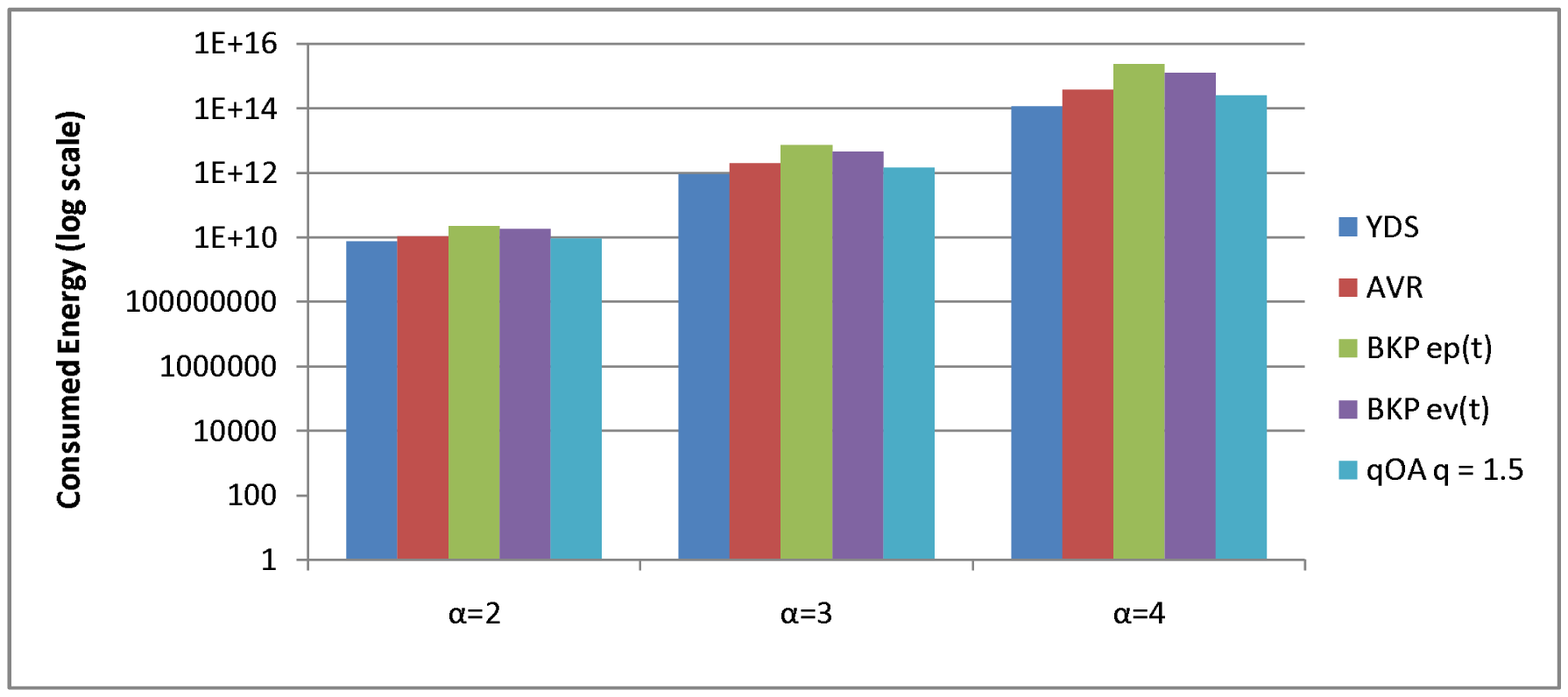, scale=0.45}
\caption{Energy consumption for a flat workload.}
\label{fig:energy-flat}
\end{figure}

\begin{figure}
\centering
\epsfig{file=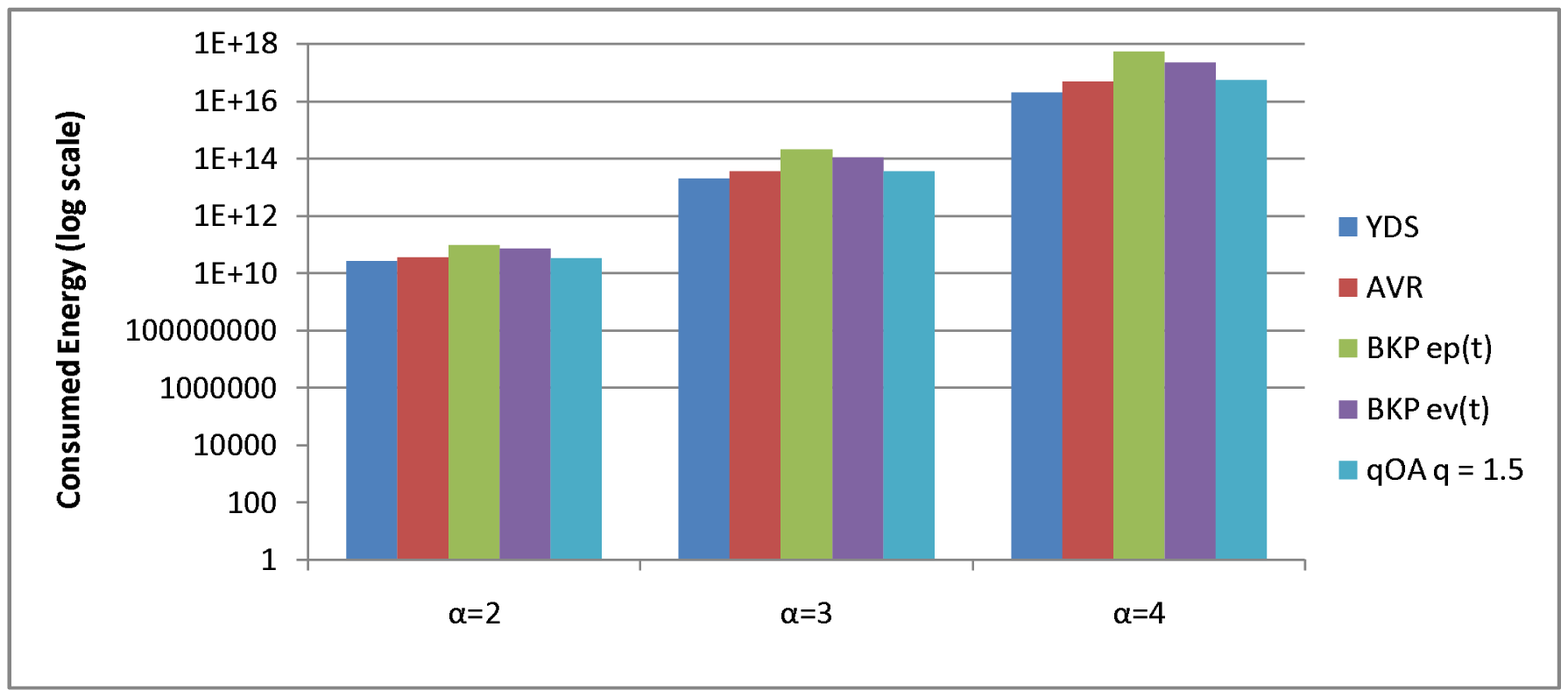, scale=0.45}
\caption{Energy consumption for a fixed span workload.}
\label{fig:energy-fixed}
\end{figure}

\begin{figure}
\centering
\epsfig{file=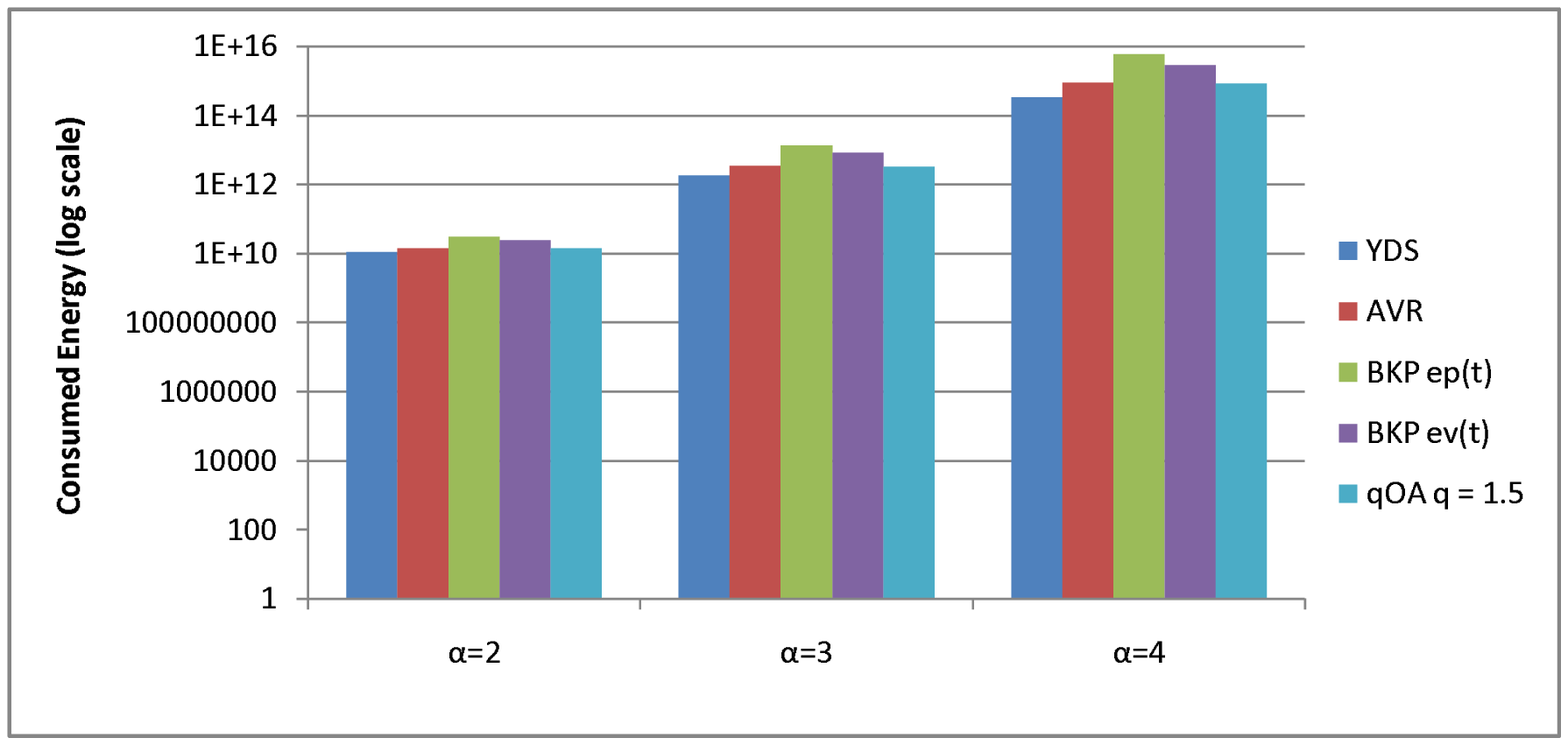, scale=0.45}
\caption{Energy consumption for a moderately spiky workload.}
\label{fig:energy-moderate}
\end{figure}

\begin{figure}
\centering
\epsfig{file=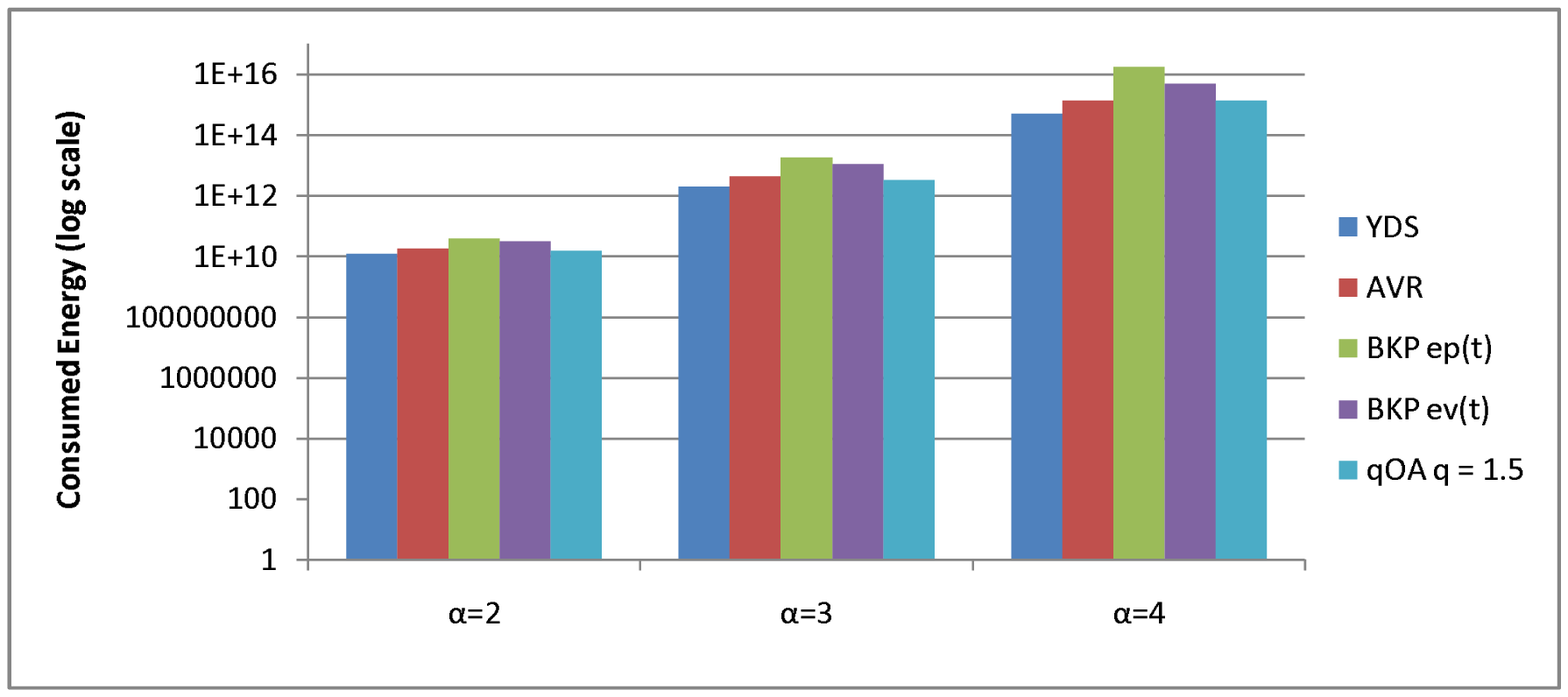, scale=0.45}
\caption{Energy consumption for a highly spiky workload.}
\label{fig:energy-high}
\end{figure}

Now that the choice of $q$ has been addressed, we can compare
the different algorithms.
In our experiments, the schedule produced by $\qOA$ (with optimal $q$)
always consumed less energy than the schedules produced by 
$\AVR$ or $\BKP$.
In fact, $\BKP$ consistently used the most energy of the algorithms we
compared.
Figures \ref{fig:energy-flat}--\ref{fig:energy-high} show the energy
consumed by each algorithm's schedule on typical instances of each
type of workload.
Since the competitive ratio of $\BKP$ improves relative to the
competitive ratio of the other algorithms as $\alpha$ increases,
we tried values of $\alpha$ up to 12 and found
that $\qOA$ still always used less energy than $\BKP$.

\begin{figure}
\centering
\epsfig{file=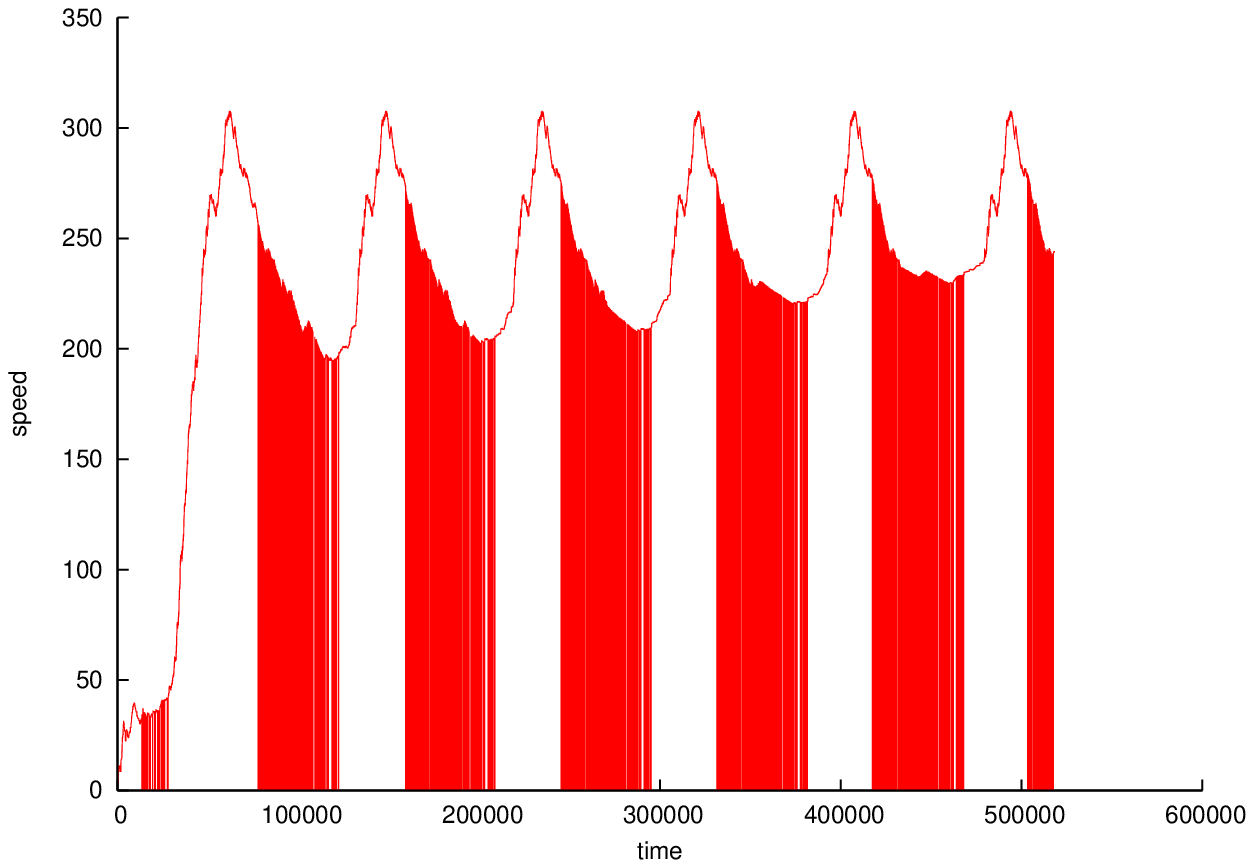, scale=.6}
\caption{Plot of $\BKP$ $ev(t)$ schedule for a flat workload.}
\label{fig:BKP_vt_Flat}
\end{figure}

\begin{figure}
\centering
\epsfig{file=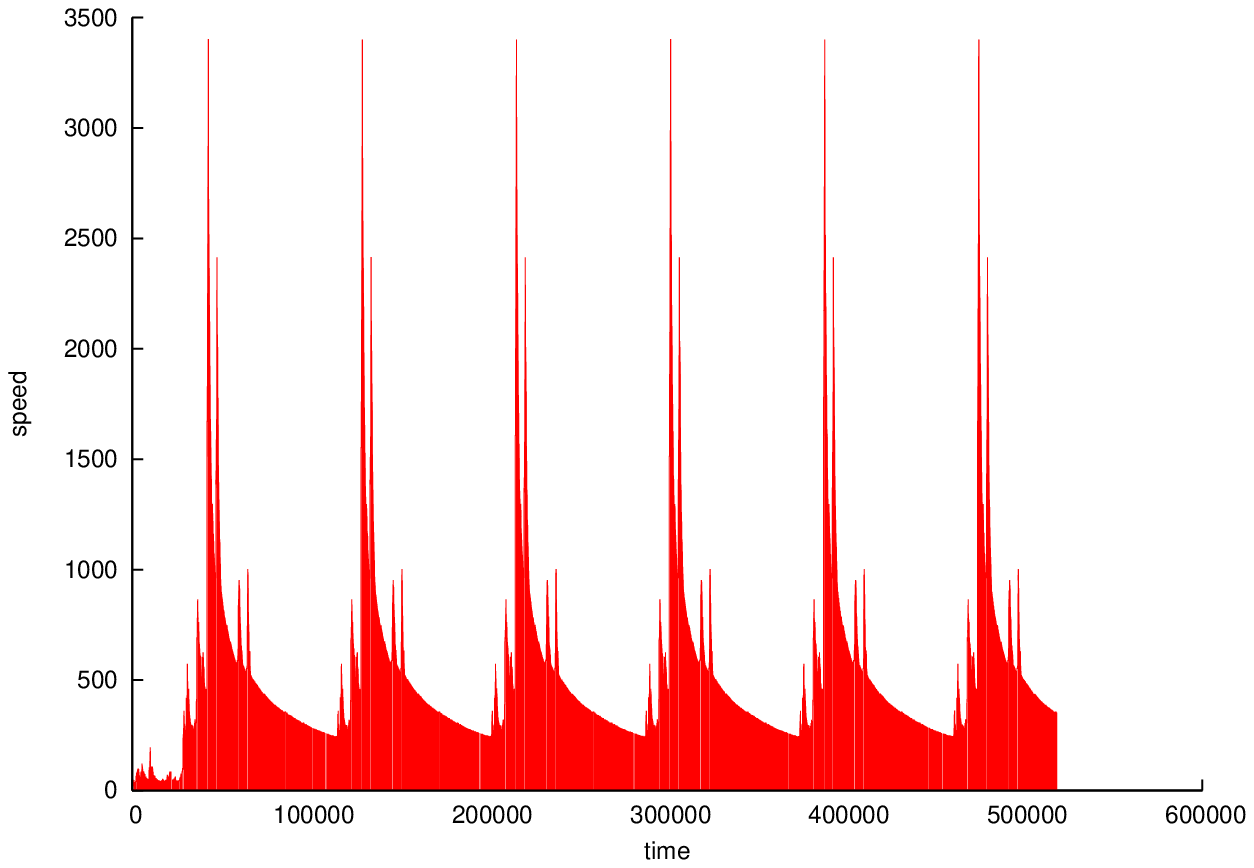, scale=.6}
\caption{Plot of $\BKP$ $ev(t)$ schedule for a fixed span workload.}
\label{fig:BKP_vt_Fixed}
\end{figure}

\begin{figure}
\centering
\epsfig{file=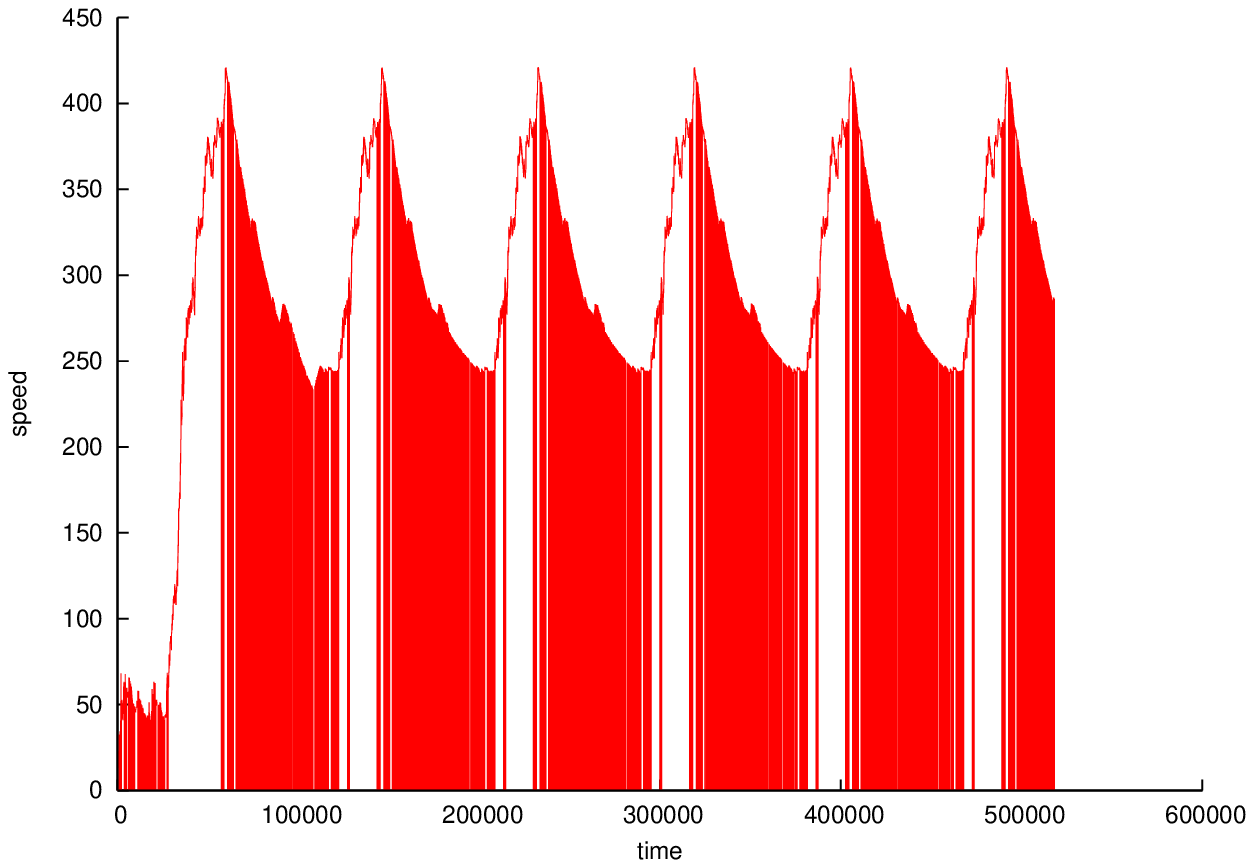, scale=.6}
\caption{Plot of $\BKP$ $ev(t)$ schedule for a moderately spiky workload.}
\label{fig:BKP_vt_Moderate}
\end{figure}

\begin{figure}
\centering
\epsfig{file=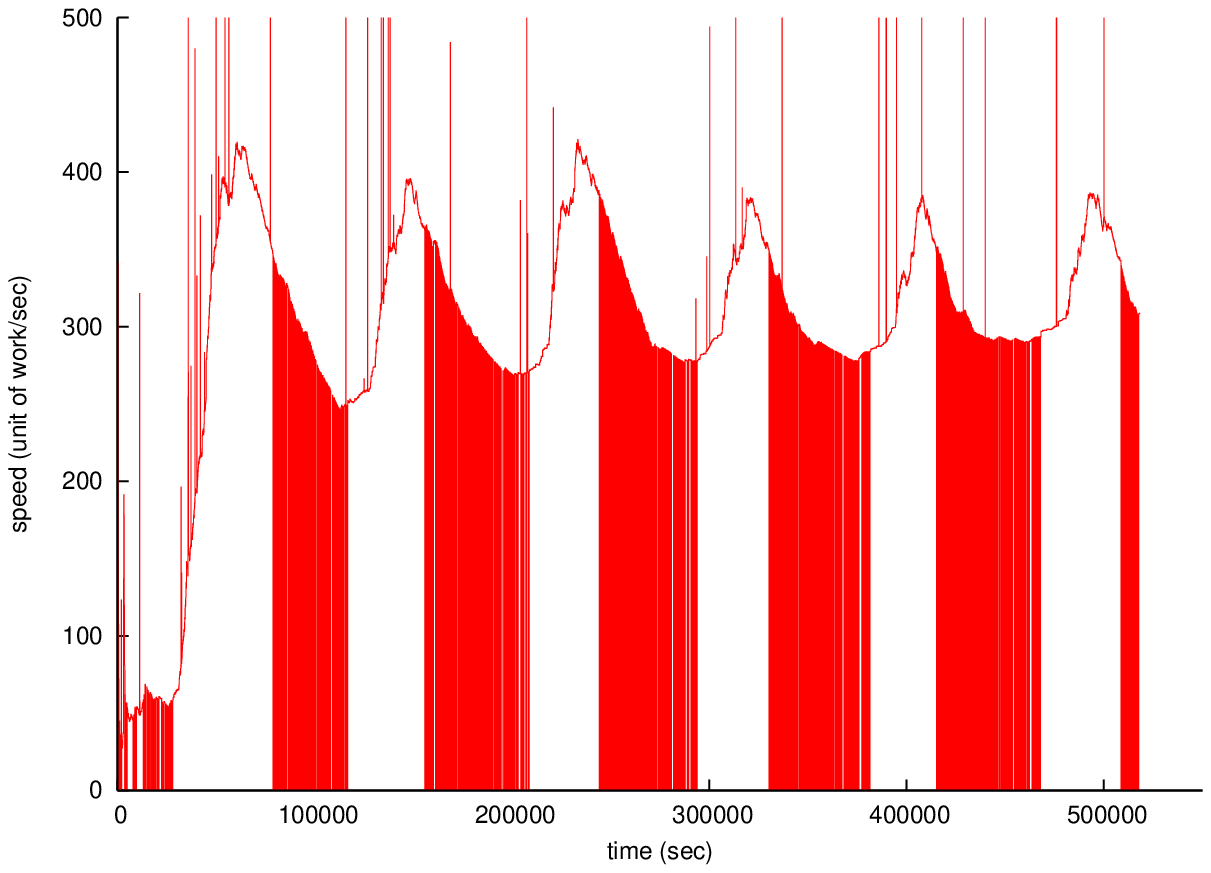, scale=.6}
\caption{Plot of $\BKP$ $ev(t)$ schedule for a highly spiky workload.}
\label{fig:BKP_vt_High}
\end{figure}

What causes the relatively poor performance of $\BKP$?
For our inputs, it seems to consistently choose a high speed at
which to run. $\BKP$ needs to use a 
high multiplicative factor $e$ times its current estimated load in order to guarantee a feasible schedule.
In addition, $\BKP$'s calculated speed can be increased by
jobs that have already finished.
Observe the $\BKP$ schedules depicted in Figures \ref{fig:BKP_vt_Flat}--\ref{fig:BKP_vt_High}.
The area under the curve appears partially filled because the
algorithm keeps switching between a high speed and being idle
(i.e. running at speed 0) because it has finished all released jobs.
Comparison with Figures \ref{fig:YDS_Flat}--\ref{fig:YDS_High}, which
show the optimal schedule for the same workloads, confirms that $\BKP$
does indeed use much higher speeds than necessary.

\begin{figure}
\centering
\epsfig{file=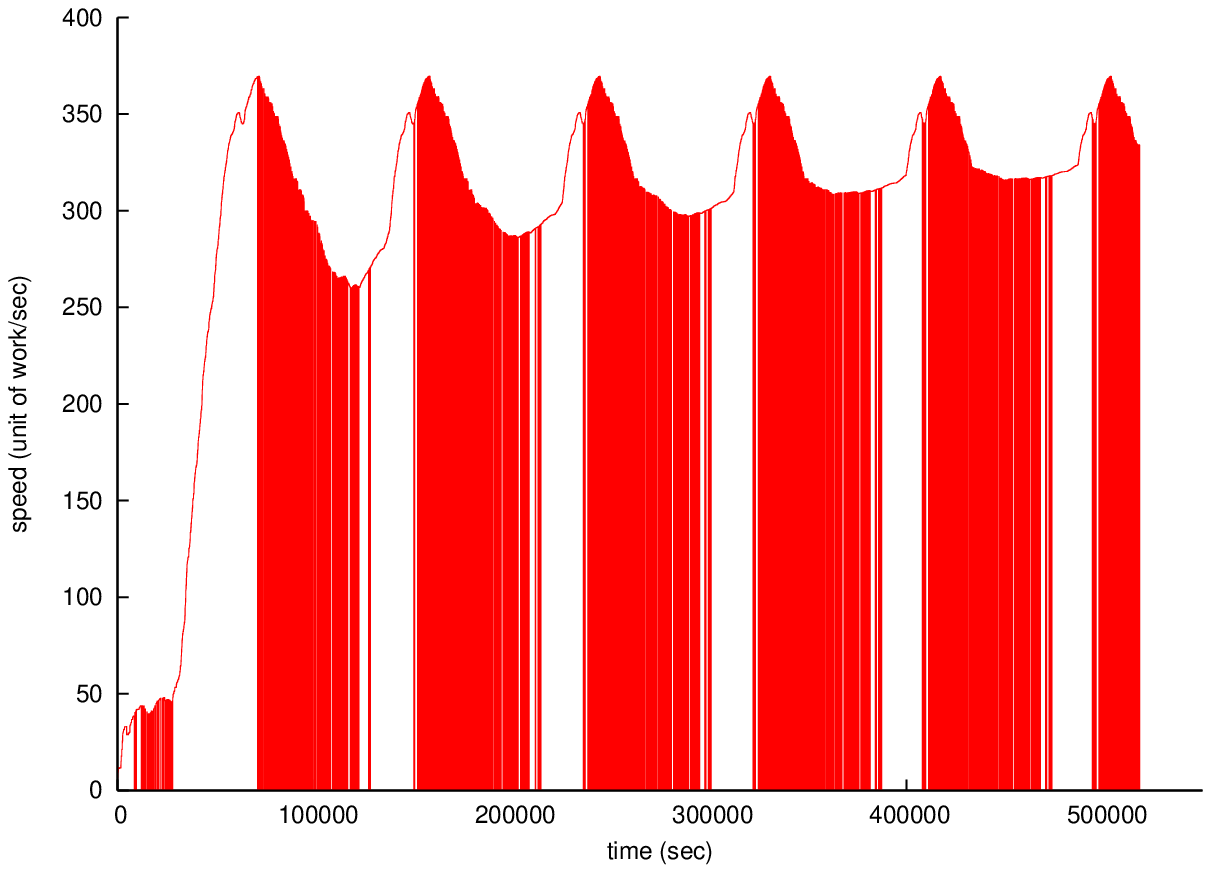, scale=.6}
\caption{Plot of $\BKP$ $ep(t)$ schedule for a flat workload.}
\label{fig:BKP_pt_Flat}
\end{figure}

Comparing the energy consumption of $\BKP$ schedules when speed is
computed using $ep(t)$ and $ev(t)$, we notice that less energy is
consumed when speed is computed using $ev(t)$.
This is demonstrated in
Figures \ref{fig:energy-flat}--\ref{fig:energy-high}.
To see why this occurs, compare Figures \ref{fig:BKP_vt_Flat} and
\ref{fig:BKP_pt_Flat}, which respectively give the 
$\BKP$ schedules using $ev(t)$ and $ep(t)$ for the flat workload.
Notice that the schedule using $ep(t)$ has higher peaks.
This is not surprising given that it seems that both versions of $\BKP$  are
running too fast at critical times, and we know that $v(t) \le p(t)$.

Our next several observations concern the optimal value of $q$ for
algorithm $\qOA$ on different workloads.
As noted above, the performance of $\qOA$ is not highly
sensitive to the exact value of $q$, but its value nonetheless
does matter.
Our explanation for why different workloads favor different values of
$q$ focuses on the spikes in schedule speed.
These spikes have disproportionate affect on the energy consumption because
raising speed to $\alpha$ causes the power to be much higher at
these times due to the convexity of the power function.
The spikes occur because the workload itself contains periods when more work
arrives, but their height is affected by two factors related to the
value of $q$.
The first factor relates to the amount of work arriving before the spike
that must be finished during it.
A higher value of $q$ tends to reduce the amount of this type of work
because higher $q$ causes $\qOA$ to
run faster before the spike, thereby reducing the optimal speed during
the spike.
The second factor, which works against the first, is that $\qOA$ runs
at $q$ times the optimal speed, including during the spike.
Thus, a large value of $q$ may increase the speed of a spike even if
the optimal speed during that time has been reduced.
We believe that the optimal value of $q$ for different types of
workloads is largely explained by the interaction of these factors on
each type of workload. 

\begin{figure}
\centering
\epsfig{file=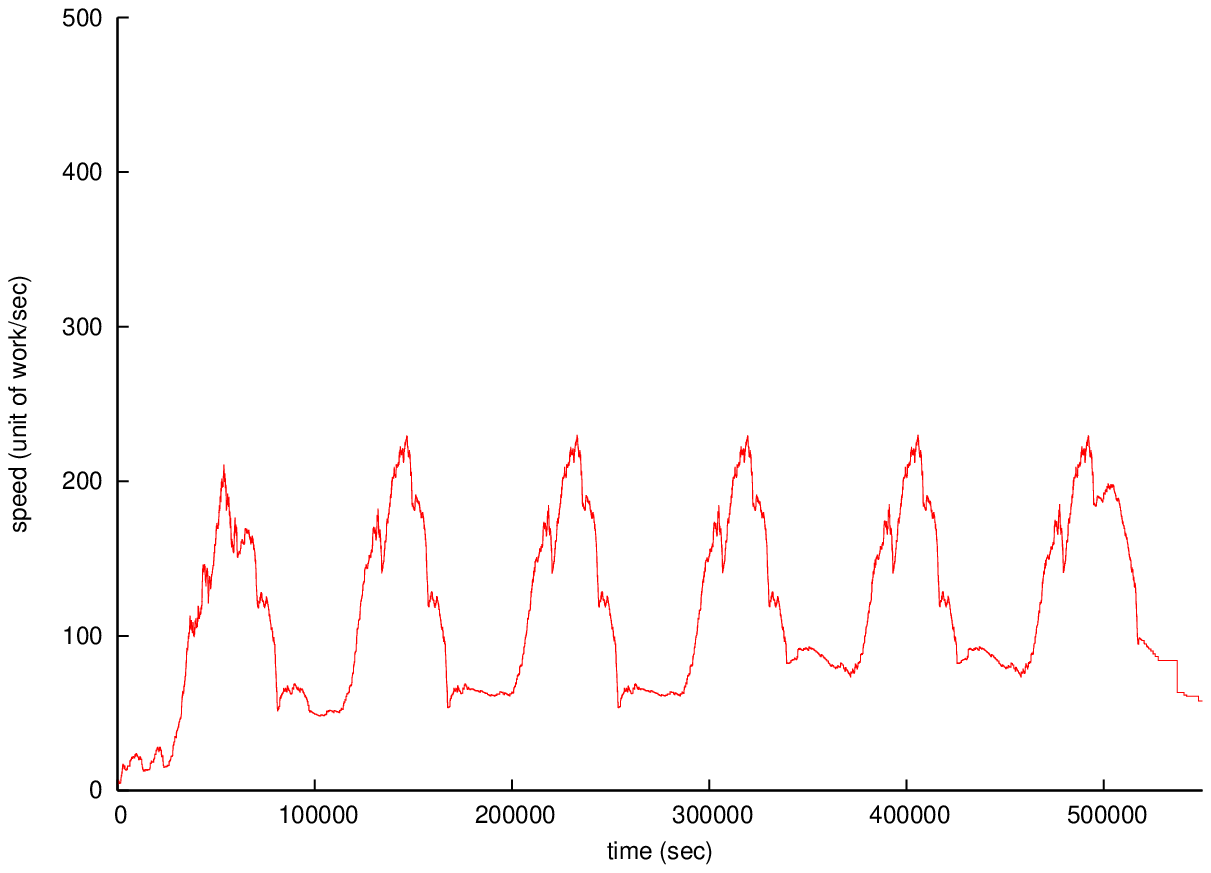, scale=.6}
\caption{Plot of $\qOA$ schedule with $q = 1.5$ for a flat workload.}
\label{fig:qOA_Flat}
\end{figure}

\begin{figure}
\centering
\epsfig{file=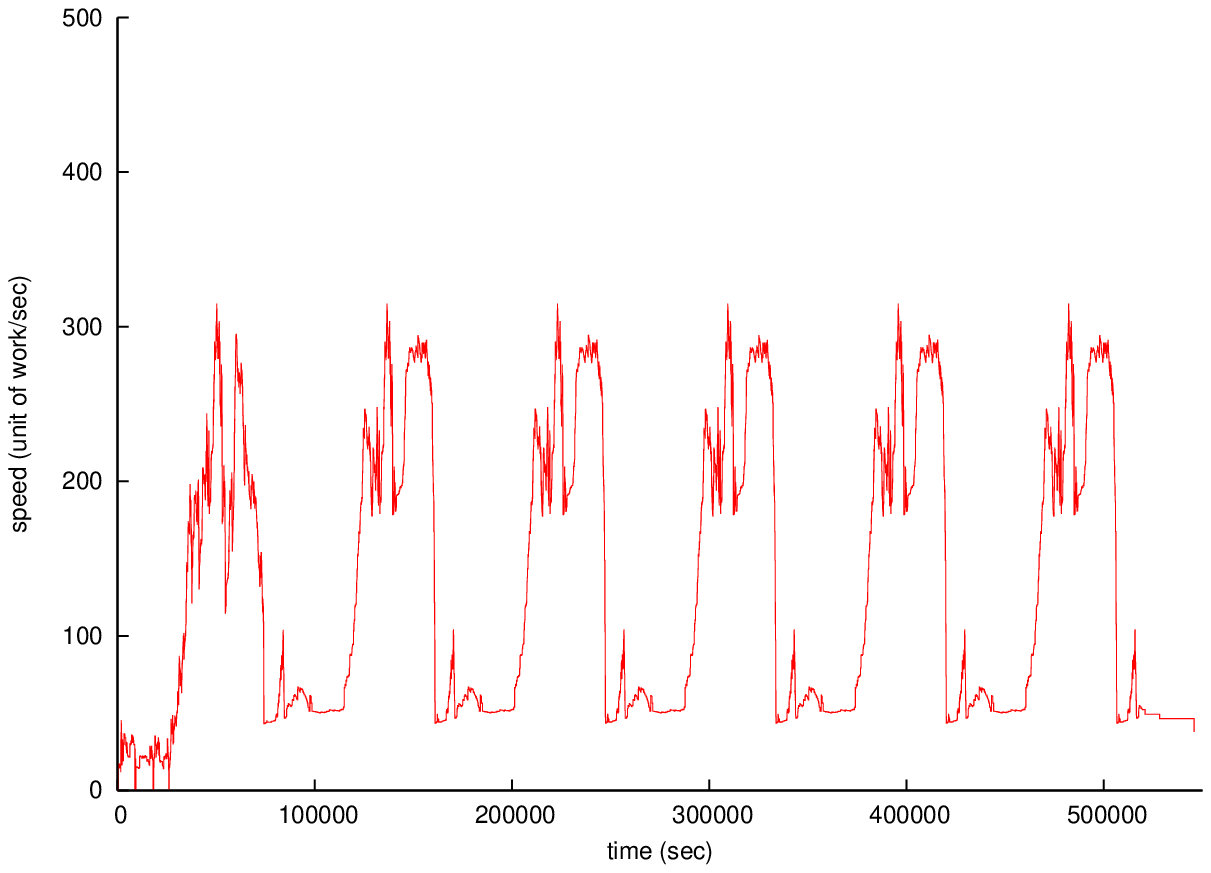, scale=.6}
\caption{Plot of $\qOA$ schedule with $q = 1.5$ for a moderately
  spiky workload.}
\label{fig:qOA_Moderate}
\end{figure}

First consider flat and moderately spiky workloads.
For these workloads, the optimal value of $q$ is usually high---
typically around 4 or higher.
It is generally higher for moderately spiky workloads than for flat
ones, as demonstrated in Figures \ref{fig:energy-flat} and
\ref{fig:energy-moderate}.
To explain these observations, we refer back to the optimal schedules
for these workloads shown in Figures \ref{fig:YDS_Flat} and
\ref{fig:YDS_Moderate}.
These figures show that the spikes in the optimal schedule are fairly
broad, with the optimal schedule for the flat workload exhibiting
smaller spikes than the optimal schedule of the moderately spiky
workload.
The broad spikes allow the benefits of higher $q$ to be felt since
finishing work early creates a narrower (but taller) spike.
The difference between the workloads occurs because 
the flat workload, where the arrival rate of work varies less (smaller
spikes in the optimal schedule), does not benefit as much from
increasing $q$.
The $\qOA$ schedules corresponding to the optimal schedules depicted in
Figures \ref{fig:YDS_Flat} and
\ref{fig:YDS_Moderate} are shown in 
Figures \ref{fig:qOA_Flat} and \ref{fig:qOA_Moderate}.

As occurs in Figures \ref{fig:energy-flat} and
\ref{fig:energy-moderate}, we observed that optimal $q$ usually
increases with increasing $\alpha$ for both flat and
moderately spiky workloads.
Increasing $\alpha$ raises the penalty for having spikes in the
schedule, so the workloads benefit from a slightly higher $q$, which
finishes work slightly earlier and shrinks the spikes.

\begin{figure}
\centering
\epsfig{file=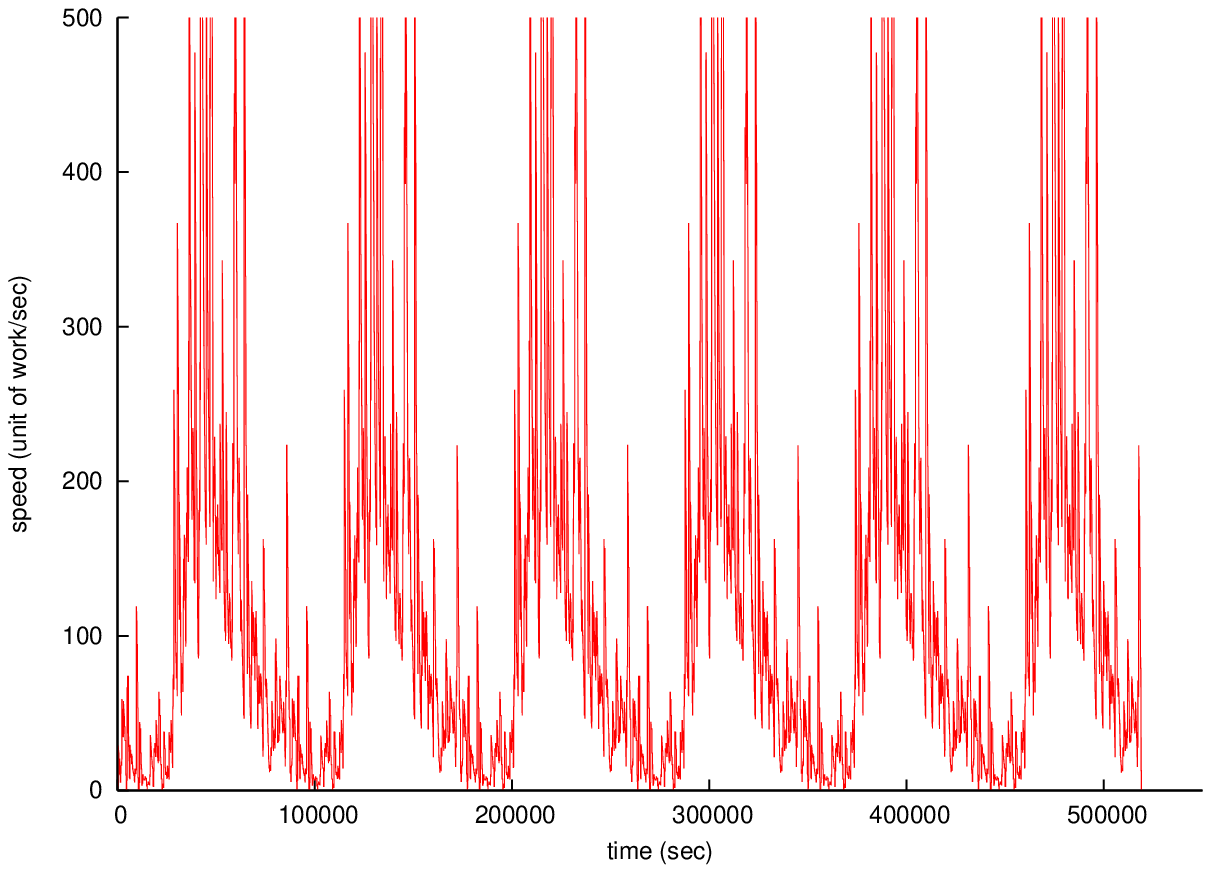, scale=.6}
\caption{Plot of $\qOA$ schedule with $q = 1.5$ for a fixed span
  workload.}
\label{fig:qOA_Fixed}
\end{figure}

\begin{figure}
\centering
\epsfig{file=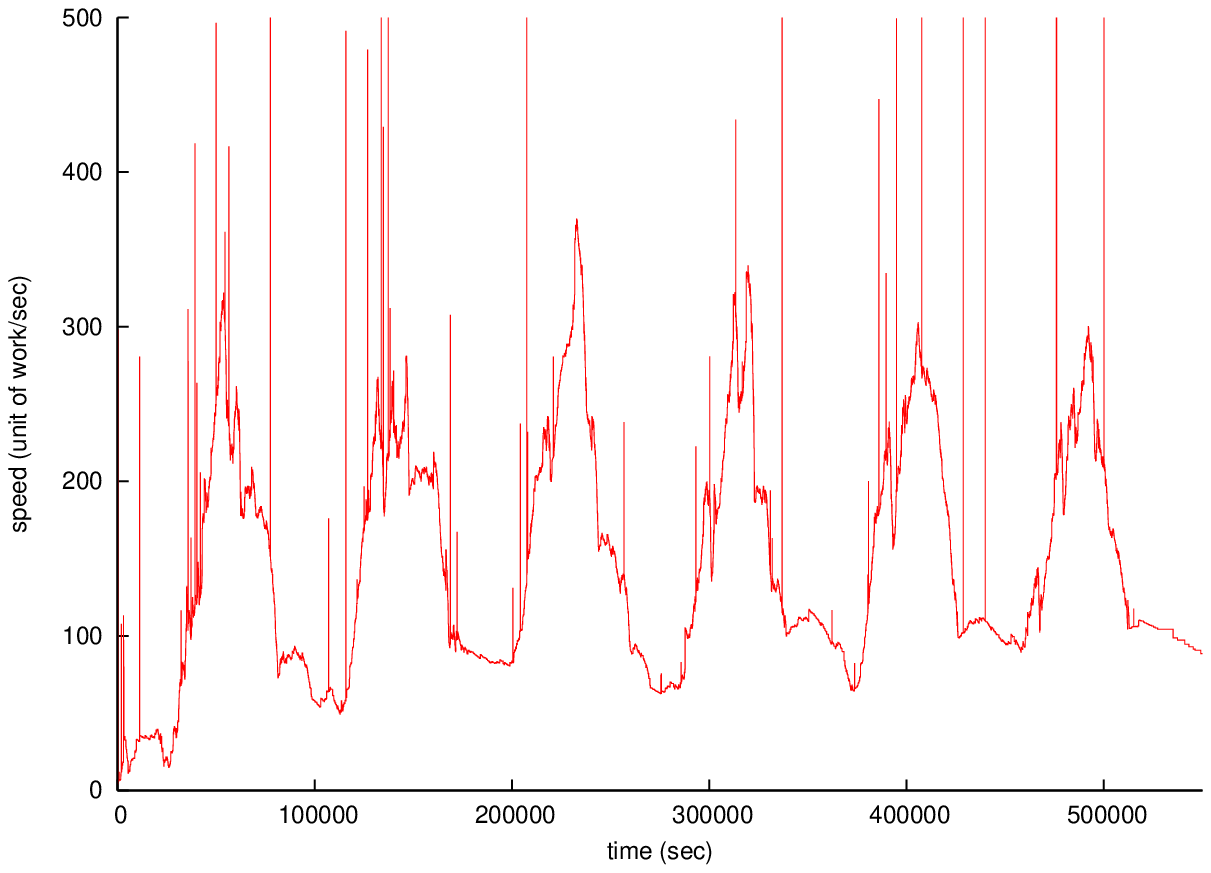, scale=.6}
\caption{Plot of $\qOA$ schedule with $q = 1.5$ for a highly spiky
  workload.}
\label{fig:qOA_High}
\end{figure}

Now we turn our attention to the value of $q$ in fixed span and
highly spiky workloads.
For these workloads, optimal $q$
is usually very low (near 1), as demonstrated in Figures
\ref{fig:energy-fixed} and \ref{fig:energy-high}.
To explain this, we again examine the optimal schedules for these
workloads; see Figures \ref{fig:YDS_Fixed} and \ref{fig:YDS_High}.
These optimal schedules have a number of very tall, very narrow spikes,
indicating the arrival of a large amount of urgent work.
The narrowness of the spike decreases the benefit of
increasing $q$ because the schedule quickly runs out of urgent work.
The height of the spike also increases the cost of large $q$ because
running at a greater multiple of the optimal speed makes the tall
spikes even taller.
The $\qOA$ schedules corresponding to the optimal schedules depicted in
Figures \ref{fig:YDS_Fixed} and \ref{fig:YDS_High} are shown in 
Figures \ref{fig:qOA_Fixed} and \ref{fig:qOA_High}.

As $\alpha$ increases, the optimal value of $q$ 
for fixed span and highly spiky workloads decreased in our
experiments (as demonstrated in Figures 
\ref{fig:energy-fixed} and \ref{fig:energy-high}), which is consistent
with the observation that increasing $\alpha$ increases the penalty
for tall spikes.

Regarding the fixed span workload, the observation that optimal
$q$ is near 1 holds for all the span lengths we tried.
We offer a partial justification for this by discussing the extreme cases.
With a short fixed span, there is little overlap between jobs and
little reason to finish one before the next arrives since they are
largely independent.
With a long fixed span, there is a lot of overlap between the
spans of jobs, allowing the optimal algorithm enough knowledge
to find a good schedule, which then does not benefit by a speed increase.
In both cases, the best $q$ will tend to be low.

\begin{figure}
\centering
\epsfig{file=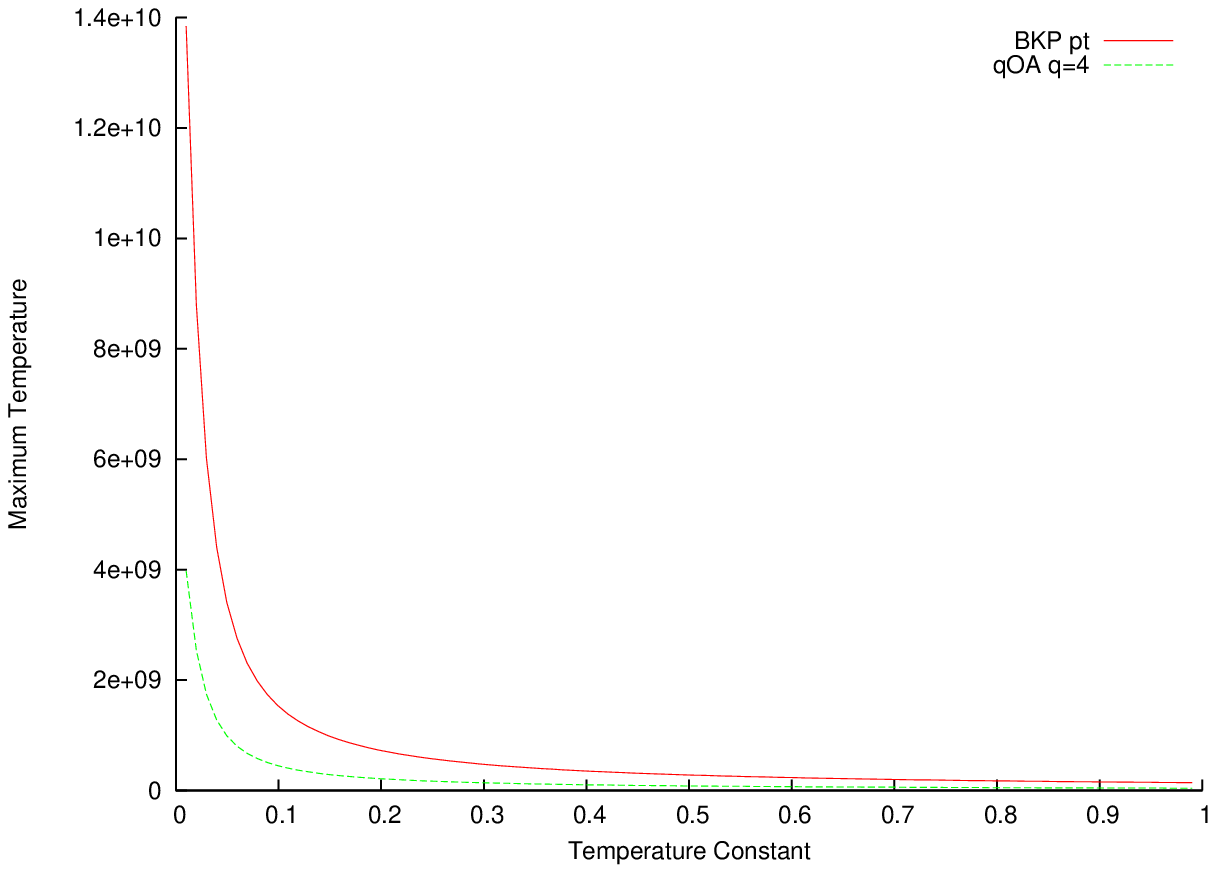, scale=.6}
\caption{Plot of maximum temperature vs. cooling parameter for schedules of $\BKP$ $ev(t)$ and $\qOA$ with $q=1.5$, for a moderately spiky workload using $\alpha=3$}
\label{fig:temperature_plot}
\end{figure}

In addition to comparing the algorithms with respect to energy
consumption, we also compare them with respect to the maximum
temperature reached by the schedules they compute.
Our temperature calculations use a discrete approximation.
We considered a range of values for the parameter $b$, and a time step of 0.1
seconds for the discrete approximation.
We compared the algorithms using the parameters: $\alpha=2, 3, 4$,
and a wide range of cooling parameters $b$.
We found that the
performance of the algorithms relative to each other with respect to
temperature is the same as their relative performance with respect to
energy consumption; from best to worst, the order was $\YDS$, $\qOA$,
$\AVR$, $\BKP$ using $ev(t)$, and $\BKP$ using $ep(t)$.
We did observe that the optimal value of $q$ for $\qOA$ could be
slightly different for minimizing temperature than for minimizing
energy consumption.
(Although none of the algorithms take $\alpha$ into account when
calculating speed, variations in $\alpha$ do favor differently shaped
schedules.)
Figure \ref{fig:temperature_plot} plots maximum
temperature as a function of the cooling parameter for the schedules
produced by $\BKP$ $ev(t)$ and $\qOA$ ($q=1.5$) for a moderately
spiky workload with $\alpha=3$.

\section{Conclusion}

In summary, if you order the 
candidate speed scaling algorithms in the 
literature by the best competitive ratio
that has been proved, this is exactly the order that these
algorithms finished in our experimental horse race.
We performed many more experiments than the 
representative sample that we report here,
and we saw the same ordering of the algorithms
across a wide range of different input distributions.
So we don't believe that these experimental results
are due to any particularities in the input distributions that used.
We thus believe that these experimental results 
can be viewed as a victory for competitive analysis as
a predictor of experimental performance, even though that
is the not the main goal of competitive analysis.

\bibliographystyle{plain}
\bibliography{alenex}

\begin{thebibliography}{10}

\bibitem{AMS}
S.~Albers, F.~M\"{u}ller, and S.~Schmelzer.
\newblock Speed scaling on parallel processors.
\newblock In {\em Proc. ACM Symposium on Parallel Algorithms and Architectures
  (SPAA)}, pages 289--298, 2007.

\bibitem{BBCP}
N.~Bansal, D.P. Bunde, H.-L. Chan, and K.~Pruhs.
\newblock Average rate speed scaling.
\newblock In {\em Latin American Theoretical Informatics Symposium}, 2008.

\bibitem{qOA}
N.~Bansal, H.-L. Chan, and K.~Pruhs.
\newblock Improved bounds for speed scaling in devices obeying the cube-root
  rule.
\newblock In {\em SODA}, 2009.
\newblock submitted.

\bibitem{BKP}
N.~Bansal, T.~Kimbrel, and K.~Pruhs.
\newblock Speed scaling to manage energy and temperature.
\newblock {\em JACM}, 54(1), 2007.

\bibitem{BKPSTACS}
N.~Bansal and K.~Pruhs.
\newblock Speed scaling to manage temperature.
\newblock In {\em STACS}, pages 460--471, 2005.

\bibitem{CCL+}
H.-L. Chan, W.-T. Chan, T.-W. Lam, L.-K. Lee, K.-S. Mak, and P.W.~H. Wong.
\newblock Energy efficient online deadline scheduling.
\newblock In {\em SODA '07: Proceedings of the eighteenth annual ACM-SIAM
  symposium on Discrete algorithms}, pages 795--804, 2007.

\bibitem{KK}
W.-C. Kwon and T.~Kim.
\newblock Optimal voltage allocation techniques for dynamically variable
  voltage processors.
\newblock In {\em Proc. ACM-IEEE Design Automation Conf.}, pages 125--130,
  2003.

\bibitem{LLY}
M.~Li, B.J. Liu, and F.F. Yao.
\newblock Min-energy voltage allocation for tree-structured tasks.
\newblock {\em Journal of Combinatorial Optimization}, 11(3):305--319, 2006.

\bibitem{LY}
M.~Li and F.F. Yao.
\newblock An efficient algorithm for computing optimal discrete voltage
  schedules.
\newblock {\em SIAM J. on Computing}, 35:658--671, 2005.

\bibitem{YDS}
F.~Yao, A.~Demers, and S.~Shenker.
\newblock A scheduling model for reduced {CPU} energy.
\newblock In {\em Proc. IEEE Symp. Foundations of Computer Science}, pages
  374--382, 1995.

\end{thebibliography}

\end{document}